\documentclass[floats,floatfix,showpacs,amssymb,prd,twocolumn,superscriptaddress,nofootinbib]{revtex4-1}
\usepackage{adjustbox}
\usepackage{anyfontsize}
\usepackage{amssymb}
\usepackage{amsmath}
\usepackage{amsfonts}
\usepackage{amsbsy}
\usepackage{txfonts}

\usepackage{subfigure}
\usepackage{gensymb}
\usepackage{soul}
\usepackage{wrapfig}

\usepackage[T1]{fontenc}

\def\namedlabel#1#2{\begingroup
    #2%
    \def\@currentlabel{#2}%
    \phantomsection\label{#1}\endgroup
}

\usepackage{mathrsfs}

\usepackage{bm}

\newcommand{\mnras}{Mon. Not. R. Astron. Soc.}

\newcommand{\aaps}{Astron. \& Astrophys. Suppl.}

\newcommand{\dG}{\Delta G/G_\text{N}}

\newcommand{\mv}{M_\text{vir}}
\newcommand{\rv}{R_\text{vir}}

\newcommand{\lc}{\lambda_C}

\DeclareMathOperator{\Ei}{Ei}
\DeclareMathOperator{\erf}{erf}

\usepackage[utf8]{inputenc}

\usepackage{graphicx}
\usepackage{epsfig}
\usepackage{epstopdf}

\usepackage[usenames,dvipsnames]{xcolor}

\usepackage{verbatim,mathtools,needspace,enumitem,etoolbox}

\definecolor{linkcolor}{rgb}{0.0,0.3,0.5}

\usepackage[unicode, colorlinks=true, linkcolor=linkcolor, citecolor=linkcolor, filecolor=linkcolor,urlcolor=linkcolor, pdfusetitle]{hyperref}

\usepackage{epstopdf}

\usepackage{bm}
\usepackage{dcolumn}

\usepackage{longtable}
 \usepackage[normalem]{ulem}

\definecolor{romared}{RGB}{142,0,28}
\hypersetup{colorlinks=true,
            citecolor=romared,
            linkcolor=romared,
            urlcolor=romared}
\begin{document}

\title{Galaxy morphology rules out astrophysically relevant Hu-Sawicki $f(R)$ gravity}

\author{Harry Desmond}
\email{harry.desmond@physics.ox.ac.uk}
\affiliation{Astrophysics, University of Oxford, Denys Wilkinson Building, Keble Road, Oxford OX1 3RH, UK}

\author{Pedro G. Ferreira}
\affiliation{Astrophysics, University of Oxford, Denys Wilkinson Building, Keble Road, Oxford OX1 3RH, UK}

\raggedbottom

\begin{abstract}
$f(R)$ is a paradigmatic modified gravity theory that typifies extensions to General Relativity with new light degrees of freedom and hence screened fifth forces between masses. These forces produce observable signatures in galaxy morphology, caused by a violation of the weak equivalence principle due to a differential impact of screening among galaxies' mass components. We compile statistical datasets of two morphological indicators --  offsets between stars and gas in galaxies and warping of stellar disks -- and use them to constrain the strength and range of a thin-shell-screened fifth force. This is achieved by applying a comprehensive set of upgrades to past work \cite{HIOC,warp}: we construct a robust galaxy-by-galaxy Bayesian forward model for the morphological signals, including full propagation of uncertainties in the input quantities and marginalisation over an empirical model describing astrophysical noise. Employing more stringent data quality cuts than previously we find no evidence for a screened fifth force of any strength $\dG$ in the Compton wavelength range $0.3-8$ Mpc, setting a $1\sigma$ bound of $\dG<0.8$ at $\lambda_C=0.3$ Mpc that strengthens to $\dG<3\times10^{-5}$ at $\lambda_C=8$ Mpc. These are the tightest bounds to date beyond the Solar System by over an order of magnitude. For the Hu-Sawicki model of $f(R)$ with $n=1$ we require a background scalar field value $f_{R0} < 1.4 \times 10^{-8}$, forcing practically all astrophysical objects to be screened. We conclude that this model can have no relevance to astrophysics or cosmology.
\end{abstract}

\date{\today}

\maketitle

\section{Introduction}
\label{sec:intro}

With cosmic acceleration unexplained, significant tensions between $\Lambda$CDM and observations emerging and large regions of gravitational parameter space still unexplored, modified gravity theories retain appeal. On astrophysical scales the search for new gravitational physics is largely tantamount to the search for additional fundamental forces. Such \emph{fifth forces} arise generically when a new field couples to matter, as happens in almost all extensions to the Einstein--Hilbert or Standard Model actions \cite{Clifton_rev}. To prevent violation of precise constraints on gravity within the Solar System, fifth forces must be \emph{screened}, so that their range or coupling to matter depends on the local gravitational environment. Screening mechanisms fall into two main categories: those that operate by the thin-shell mechanism and are governed by the gravitational potential (e.g. chameleon \cite{chameleon_prl, chameleon_prd}, symmetron \cite{symmetron} and dilaton \cite{dilaton}) and those that operate kinetically and are governed by derivatives of the potential (e.g. K-mouflage \cite{kmouflage} and Vainshtein \cite{vainshtein}). Reviews of screened modified gravity can be found in \cite{Jain_rev, Joyce_rev, Khoury_rev, NPP}.

The observational consequence of a screened fifth force in astrophysics is a violation of the weak equivalence principle, of a nature depending on the relative coupling of the new field to different types of object. We focus here on thin-shell screening, as our specific emphasis will be $f(R)$ gravity which screens by the chameleon mechanism \cite{Carroll, f(R)_rev_1, f(R)_rev_2}. Under this type of screening, main-sequence and denser stars in otherwise unscreened galaxies may self-screen due to their relatively deep gravitational potential. This causes two particularly interesting observable signatures in galaxy morphology \cite{Jain_Vanderplas, Vikram, HIOC, warp}: i) a separation of stars and gas as the latter is pulled in the direction of an external fifth-force field but not the former, and ii) a warping of stellar disks due to the potential gradient arising from their offsets from the halo centres.

The aim of this paper is to use these signals to set constraints on the strength and range of thin-shell-screened fifth forces, with particular emphasis on the Hu-Sawicki model of $f(R)$ gravity \cite{Hu_Sawicki}. Building on previous work \cite{maps, letter, HIOC, warp}, we develop a Monte-Carlo-based technique in the framework of Bayesian statistics to forward-model the signals on a galaxy-by-galaxy basis in a large sample ($\mathcal{O}(10^4)$) of galaxies. This works by combining models for the gravitational and fifth-force fields in the local universe and the internal structures of the galaxies, and marginalising over uncertainties in the inputs as well as additional noise parameters describing astrophysical (i.e. non-fifth-force) contributions to the observables. The predictions are then compared via a Markov Chain Monte Carlo algorithm with the measured values from photometric optical and HI surveys to constrain the fifth-force strength relative to gravity, $\dG$, and range, $\lambda_C$, which corresponds to the Compton wavelength of the new field. Relative to our previous work, we use deeper screening maps based on additional data treated more homogeneously, model the fifth force more realistically, calibrate the astrophysical noise models empirically, enlarge the test datasets and apply machine learning techniques to model unresolved fifth-force sources and to characterise the likelihood.

\begin{figure}
   \centering
   \includegraphics[width=0.99\linewidth]{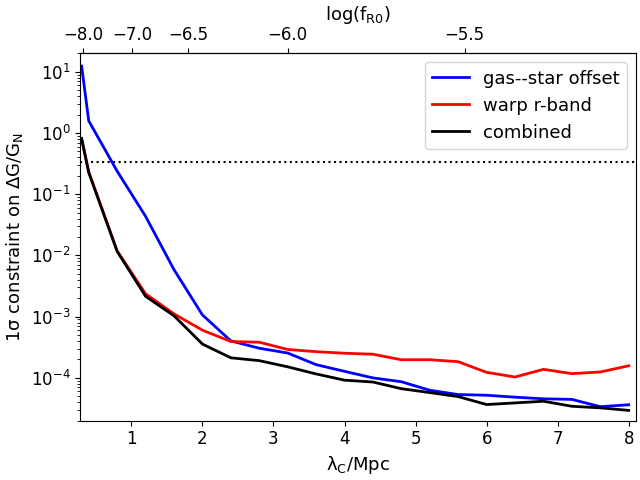}
   \caption{$1\sigma$ constraints on $\dG$ as a function of $\lambda_C$ or $f_{R0}$, for the gas--star offset (blue) and warp (red) analyses, and their combination (black). The horizontal dashed black line indicates the value of $\dG=1/3$ in $f(R)$.}
   \label{fig:constraints}
\end{figure}

Besides setting strong constraints on $\dG$, \cite{HIOC, warp} identified a region of parameter space around $\{\lambda_C = 2 \; \text{Mpc}, \dG=0.02\}$ offering significantly improvement over General Relativity (GR). We find here that this result is driven by a small number ($<10$) of galaxies with extreme measured warps and gas--star offsets in regions of unusually strong fifth-force field. On excising these potential outliers we find no significant improvement over GR for any $\lambda_C$ in the range $0.3 < \lambda_C/\text{Mpc} < 8$. We use this to set the most stringent constraints to date on thin-shell-screened fifth forces with $\mathcal{O}$(Mpc) range, which we show in Fig.~\ref{fig:constraints}. The constraint on the background scalar field value in $n=1$ Hu-Sawicki $f(R)$, in which $\dG=1/3$, is $f_{R0} < 1.4 \times 10^{-8}$. This forces practically all astrophysical objects to screen in that theory, rendering it astrophysically and cosmologically uninteresting. Strong constraints are also obtained for $n\ne1$, and we expect our conclusions to apply more broadly to viable $f(R)$ models.

The structure of this paper is as follows. In Sec.~\ref{sec:f(R)} we present $f(R)$ theory and explain the way it screens. In Sec.~\ref{sec:signals} we recap how gas--star offsets and warps come about under a thin-shell-screened fifth force, and describe the data that we use to search for them. Sec.~\ref{sec:method} details our inference method, with particular emphasis on the parts we have recently improved. Sec.~\ref{sec:results} presents our results, Sec.~\ref{sec:discussion} discusses systematic uncertainties and places our findings in the broader context, and Sec.~\ref{sec:conc} concludes. Further detail on many topics may be found in \cite{HIOC, warp}.

\section{f(R) gravity}
\label{sec:f(R)}

$f(R)$ \cite{Buchdahl, Carroll} is an archetypal modified gravity theory and a typical representative of the class of GR extensions with thin-shell-screened fifth forces. The action is found by replacing the Ricci scalar $R$ by $R+f(R)$ in the Einstein--Hilbert action, where $f$ is a general function:
\begin{equation}
S = \int d^4x \sqrt{-g} \: \frac{R+f(R)}{16 \pi G_\text{N}} + S_\text{m}
\end{equation}
with $S_\text{m}$ the matter action (we work throughout in units where $c \equiv 1$). The theory contains a dynamical scalar degree of freedom \cite{Brax_f(R)} which can be described by the derivative of the $f$ function, $f_R \equiv \frac{df}{dR}$, with equation of motion:
\begin{equation}\label{eq:f_propagation}
\nabla^2 f_R = \frac{1}{3} a^2 (\delta R - 8 \pi G_\text{N} \delta \rho).
\end{equation}
$\delta R$ and $\delta \rho$ are fluctuations in the scalar curvature and matter density around the cosmological background. The Poisson equation for the Newtonian potential is also modified:
\begin{equation}\label{eq:phi_propagation}
\nabla^2 \Phi = \frac{16 \pi G_\text{N}}{3} a^2 \delta \rho - \frac{1}{6} a^2 \delta R.
\end{equation}
These equations assume $f_R \ll 1$ and the quasistatic approximation $|\nabla f_R| \gg \partial f/\partial t$, which are always valid on subhorizon scales in cosmology.

To fully specify the dynamics of the theory the form of $f(R)$ must be fixed. Here we adopt the canonical Hu-Sawicki model \cite{Hu_Sawicki}
\begin{equation}
f(R) = -m^2 \frac{c_1 (R/m^2)^n}{1+c_2(R/m^2)^n}
\end{equation}
where $m^2 \equiv H_0^2 \Omega_m$ (at $z=0$) and $c_1$, $c_2$ and $n$ are the free parameters of the theory. Following much of the literature we choose $n=1$, although we consider alternative choices in Sec.~\ref{sec:results}. This form for $f(R)$ gives the scalar field
\begin{equation}
f_R = -\frac{c_1}{c_2} \left(\frac{m^2}{R}\right)^2.
\end{equation}
One of the remaining two parameters can be fixed by matching the $\Lambda$CDM expansion history to first order, so that $f(R) \simeq -2 \Lambda$. This requires $c_1/c_2 = 6 \Omega_\Lambda/\Omega_m$. Assuming an FRW metric the background curvature can be expressed as a function of the scale factor:
\begin{equation}\label{eq:R}
\bar{R}(a) = 3 \frac{m^2}{a^3} \left(1 + 4 \frac{\Omega_\Lambda a^3}{\Omega_m}\right),
\end{equation}
where the overbar denotes evaluation in the background. The present-day background value of the scalar field, denoted $f_{R0}$ (subscript `0' denotes evaluation at the present day and we omit the overbar for brevity), follows:
\begin{equation}
f_{R0} = -\frac{2 \Omega_\Lambda \Omega_m}{3 c_2 (\Omega_m + 4\Omega_\Lambda)^2}.
\end{equation}
Thus, given a cosmology, $f_{R0}$ is the sole degree of freedom of the theory, representing the strength of the modified gravity. GR+$\Lambda$CDM is recovered in the limit $f_{R0} \rightarrow 0$ and higher values result in greater scalar field magnitudes.

The scalar field generates a fifth force according to
\begin{equation}\label{eq:a5_0}
\vec{a}_5 = \frac{1}{2} \vec{\nabla} f_R,
\end{equation}
which in the absence of screening would act as a universal Yukawa force between masses with range given by the Compton wavelength $\lambda_C$ of the scalar field. However, the model naturally incorporates the \emph{chameleon mechanism}, whereby $\lambda_C$ depends sensitively on the ambient matter density. To see this we apply the linearised version of Eq.~\ref{eq:R} to Eq.~\ref{eq:f_propagation}, yielding:
\begin{equation}\label{eq:mKG}
\nabla^2 f_R = \frac{a^2}{\lambda_C^2} f_R - \frac{8 \pi G}{3} a^2 \delta \rho
\end{equation}
with
\begin{equation}\label{eq:lambda}
\lambda_C = \left(\frac{1}{6} \frac{\bar{R}}{f_{R0}} \left(\frac{\bar{R}}{\bar{R}_0}\right)^2\right)^{-1/2}.
\end{equation}
In the Hu-Sawicki model at cosmological densities this becomes \cite{Hu_Sawicki, Cabre}
\begin{equation}\label{eq:lambdaC_fR0}
\lambda_C \simeq 0.32 \: \sqrt{f_{R0}/10^{-8}} \: \text{Mpc}.
\end{equation}
The scalar field cannot propagate beyond the Compton wavelength, while below it Eq.~\ref{eq:phi_propagation} becomes
\begin{equation}
\nabla^2 \Phi = \frac{16 \pi G}{3} a^2 \delta \rho.
\end{equation}
This represents an enhancement of the gravitational force by a factor 4/3, equivalent to $\dG=1/3$. The linearisation around the cosmological background fails in regions of large density where $f_R \gg f_{R0}$, in which case $f_R$ is suppressed and GR is recovered. This suppression (or chameleon screening) occurs in regions of deeper gravitational potential than a threshold $\chi$, called the self-screening parameter of the theory. In the Hu-Sawicki model this is related to $\lambda_C$ and $f_{R0}$ by 
\begin{equation}\label{eq:chi_lambdaC}
\chi = \frac{3}{2} f_{R0} = \frac{3}{2} \times 10^{-8} \left(\frac{\lambda_C}{0.32 \: \text{Mpc}}\right)^2.
\end{equation}
Comprehensive reviews of $f(R)$ may be found in \cite{f(R)_rev_1, f(R)_rev_2}.

While screening is strictly a condition on the fluctuation of the gravitational potential, it has been shown in simulations \cite{Zhao_prd, Zhao_prl, Cabre} that a good approximation is achieved for real astrophysical objects by using instead the total gravitational potential sourced by masses within $\lambda_C$. Our condition for screening is therefore $|\Phi_\text{tot}| > \chi$, where $\Phi_\text{tot}$ is the sum of the object's own gravitational potential (self-screening) and that due to neighbouring objects within $\lambda_C$ (environmental screening). We will assume Eq.~\ref{eq:chi_lambdaC} to calculate $\chi$ from $\lambda_C$, although we will also consider variations in $\chi$ by up to an order of magnitude on either side of its fiducial value to simulate the effect of other $n$ values in the Hu-Sawicki model and, more generally, other thin-shell-screened theories. To further increase generality over $f(R)$ we will allow $\dG$ values other than $1/3$ in unscreened regions. This would correspond to an arbitrary coupling coefficient of the scalar field to matter.

\section{Gas--star offsets \& stellar warps}
\label{sec:signals}

\subsection{Fifth-force phenomenology}
\label{sec:signals_F5}

Consider a galaxy in a low-density environment that is sufficiently low mass not to screen itself. The diffuse gas and dark matter (DM) in this galaxy experience an acceleration, $\vec{a}_5$, in the direction of the fifth-force field sourced by surrounding unscreened mass. The main-sequence stars however are sufficiently massive that they will self-screen if $\chi \sim f_{R0} \lesssim 10^{-6}$, the Newtonian potential at their surface. This is necessary for the Sun to screen the Earth, a basic viability requirement on any thin-shell theory due to the precise tests of the equivalence principle and inverse-square law on terrestrial and Solar System scales \cite{Adelberger}. Note however that from this constraint alone a large region of parameter space remains in which dwarf galaxies in voids (with $\Phi_\text{tot} \sim 10^{-8}$) could be unscreened, which would give the theory relevance to astrophysics.

With gas and dark matter feeling a fifth force but not stars, the centroid of emission of HI (sourced by diffuse neutral hydrogen) will become displaced from that of optical emission (sourced by main-sequence stars) in the direction of the external fifth-force field. This will continue until the displacement reaches an equilibrium value $\vec{r}_*$ at which the restoring force on the stellar disk due to its offset from the halo centre exactly compensates for its not feeling the fifth force. For an enclosed mass profile $M(r)$ and external fifth-force field $\vec{a}_5$, this is given by:
\begin{equation}\label{eq:rstar}
\frac{G_\text{N} M(r_*)}{r_*^2} \: \hat{r}_* = \vec{a}_5 \: \frac{\Delta G}{G_\text{N}},
\end{equation}
where hat denotes a unit vector.

At the same time, the separation of the stellar centroid from the halo centre establishes a gravitational potential gradient across the stellar disk, causing the formation of a U-shaped warp. The equilibrium shape of the disk is set by the requirement that the total acceleration is the same at all points along it \cite{Jain_Vanderplas, warp}. This is given by
\begin{equation}\label{eq:w1_curve}
z(x) \simeq -a_{5,z} \: \frac{\Delta G}{G_\text{N}^2} \frac{|x|^3}{M(x)},
\end{equation}
where $x$ is the direction along the disk's major axis and $z$ is the perpendicular direction on the plane of the sky. Following \cite{warp} we summarise the magnitude of the warp by a dimensionless statistic
\begin{equation} \label{eq:w1_integral}
w_1 \equiv \frac{1}{(3 R_\text{eff})^3} \int_{-3R_\text{eff}}^{3R_\text{eff}} |x| \: z'(x) \: dx,
\end{equation}
where $z'(x) \equiv z(x) - \langle z \rangle$. This specifically picks out the U-shaped warps of interest here as opposed to the observationally more common S-shapes \cite{Binney_warps}. We describe in Sec.~\ref{sec:data} how $w_1$ is calculated from the data, and in Sec.~\ref{sec:model_signals} how we model it in our fifth-force inference. The key parameters used in the analysis are summarised in Table~\ref{tab:galprops}.

\begin{table*}
  \centering
  \small\addtolength{\tabcolsep}{-5pt}
      \begin{tabular}{|c|c|c|}
      \hline
    Parameter & Description & Range, reference or source\\
      \hline
      \hline
    \rule{0pt}{3.5ex}
      $\Delta G/G_\text{N}$ & Strength of unscreened fifth force relative to gravity & $\ge0$\\
    \rule{0pt}{3.5ex}
      $\lambda_C$ & Compton wavelength of scalar field = range of fifth force & $0.3-8$ Mpc\\
    \rule{0pt}{3.5ex}
      $\chi$ & Threshold Newtonian potential at which screening occurs & $1.3 \times 10^{-8} - 9.4 \times 10^{-6}$\\
    \hline
    \rule{0pt}{3.5ex}
      $\vec{r}_*$ & Displacement between optical and HI centroid on plane of sky & Secs.~\ref{sec:signals_F5} \& \ref{sec:model_signals}\\
    \rule{0pt}{3.5ex}
      $w_1$ & Magnitude of U-shaped warp & \textquotedbl \\
    \hline
    \rule{0pt}{3.5ex}
      $\Phi_\text{ext}(\vec{x}; \lambda_C)$ & Newtonian potential due to sources within $\lambda_C$ of $\vec{x}$ & Sec.~\ref{sec:gravity_maps}\\
    \rule{0pt}{3.5ex}
      $\vec{a}_5(\vec{x}; \lambda_C)$ & Fifth-force acceleration sourced by unscreened mass within $4\lambda_C$ of $\vec{x}$ & \textquotedbl\\
    \hline
    \rule{0pt}{3.5ex}
      $R_\text{eff}$ & Half-light radius of galaxy along major axis & \textit{Nasa Sloan Atlas}; Sec.~\ref{sec:model_signals}\\
    \rule{0pt}{3.5ex}
      $r_s$ & NFW scale radius of halo & Halo abundance matching; \textquotedbl\\
    \rule{0pt}{3.5ex}
      $\rho_{rs}$ & Density of NFW halo at $r=r_s$ & Halo abundance matching; \textquotedbl\\
    \rule{0pt}{3.5ex}
      $V_\text{vir}$ & Virial velocity of halo & Halo abundance matching; \textquotedbl\\
    \rule{0pt}{3.5ex}
      $\beta$ & Negative power-law slope of dark matter profile within galaxy & $0-1.5$ (0.5 fiducial); \textquotedbl\\
    \hline
    \rule{0pt}{3.5ex}
      $f_i$ & Probability that galaxy $i$ is unscreened & Sec.~\ref{sec:likelihood}\\
    \rule{0pt}{3.5ex}
      $\sigma_i$ & Width of Gaussian astrophysical noise contribution to signal for galaxy $i$ & \textquotedbl \\
    \rule{0pt}{3.5ex}
      $W_{ik}, \bar{y}_{ik}, s_{ik}$ & Weight, mean and width of Gaussian component $k$ fit to likelihood of galaxy $i$ & \textquotedbl \\
    \hline
    \end{tabular}
  \caption{Main parameters used in the paper. Groups are separated by horizontal lines, and contain i) the intrinsic parameters of the screened fifth force (that we wish to constrain), ii) the observational summary statistics with which we derive the constraints, iii) modelled properties of the galaxies' gravitational environments, iv) parameters describing the galaxies' intrinsic structures, and v) features of the likelihood function with which we perform the inference.}
  \label{tab:galprops}
\end{table*}

\subsection{Data acquisition and reduction}
\label{sec:data}

We derive a sample of gas--star offsets by cross-correlating the 100\% dataset of ALFALFA\footnote{\url{http://egg.astro.cornell.edu/alfalfa/}} \cite{Haynes_ALFALFA} with v1\_0\_1 of the NASA Sloan Atlas (NSA)\footnote{\url{https://www.sdss.org/dr13/manga/manga-target-selection/nsa/}}. ALFALFA is a blind, 7000 deg$^2$ HI survey with sensitivity to HI masses as low as $10^6 M_\odot$; the NSA is a value-added optical and near-infrared galaxy catalogue of principally Sloan Digital Sky Survey (SDSS) sources. While v1\_0\_1 of the NSA extends to $z=0.15$, there are few ALFALFA sources beyond $z=0.06$.

ALFALFA provides sky coordinates for an optical counterpart (OC) to $>98\%$ of their detections, derived by interactively cross-matching to objects in SDSS, DSS2 and the Nasa Extragalactic Database (NED) \cite{Haynes_2011}. We associate each ALFALFA source with the nearest galaxy in the NSA to the OC, provided they are within $5^{\prime\prime}$ on the plane of the sky and 10 Mpc along the line of sight. We strengthen our requirement from \cite{HIOC} on the angular separation between the HI and OC centres from $<2^\prime$ to $<1^\prime$, thus providing a more conservative cut on misassociation of the optical and radio sources. We further remove galaxies without distance (for which we use \texttt{zdist}) or S\'{e}rsic index, luminosity, radius or axis ratio information in the NSA, and ALFALFA detections with quality flag (\texttt{HIcode}) greater than 1.  Finally we impose a distance cut of 250 Mpc for reasons we describe in Sec.~\ref{sec:gravity_maps}, leaving a final sample of 15,634 galaxies. In contrast to \cite{HIOC}, ALFALFA detections not matched to an NSA source are discarded, allowing for a homogeneous analysis pipeline across the sample and eliminating a number of assumptions required when NSA information is not available. The observed $\vec{r}_*$ is simply the displacement between the centroid of HI emission and the OC on the plane of the sky, which we decompose into its RA ($r_{*,\alpha}$) and Dec ($r_{*,\delta}$) components for modelling.

Our sample of disk warps is derived entirely from the NSA. We select galaxies with S\'{e}rsic $r$-band magnitude $M_r < -17.7$, $D < 250$ Mpc, and S\'{e}rsic apparent minor-to-major axis ratio $b/a=0.15$, the minimum returned by the S\'{e}rsic profile fitting algorithm. The magnitude cut ensures that the galaxies fall within the observationally well-known part of the luminosity function, which is required to model their halo properties using halo abundance matching as described in Sec.~\ref{sec:model_signals}. The $b/a$ cut selects thin disks viewed edge-on, as assumed in our theoretical modelling of $w_1$. $r$- and $i$-band images of the resulting sample of 4,345 galaxies are then obtained from the DR16 data repository.\footnote{\url{https://data.sdss.org/sas/dr16/sdss/atlas/v1/}}

Our detailed methodology for using these images to measure $w_1$ may be found in \cite{warp} sec. IVA. Briefly, we translate and rotate the image so that the major axis is along $x$ and the intensity-weighted mean position of the disk integrated along $x$ is at $z=0$. We then measure the intensity-weighted mean position $\bar{z}'(x)$ in the $z$ direction at each pixel in $x$ between $-3 R_\text{eff} < x < 3 R_\text{eff}$ and $-3 R_\text{eff} \: b/a < z < 3 R_\text{eff} \: b/a$. Finally we approximate the integral of Eq.~\ref{eq:w1_integral} by a sum over pixels, and calculate $w_1$ according to
\begin{equation}\label{eq:w1_obs}
w_1 = \frac{1}{(3R_\text{eff})^3} \sum_{x=-3R_\text{eff}}^{x=3R_\text{eff}} |x| \: \bar{z}'(x).
\end{equation}
To ensure that the disks are sufficiently regular for the approximations used to calculate the fifth-force expectation for $w_1$ to apply (see Sec.~\ref{sec:model_signals}), and to remove outliers, we cut galaxies with $|w_1|>0.002$. This leaves a final warp sample size of 4,149.

\begin{figure*}
  \centering
  \includegraphics[width=0.49\textwidth]{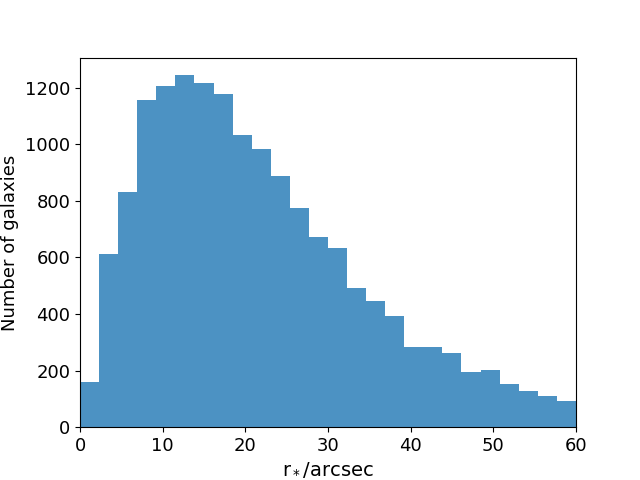}
  \includegraphics[width=0.49\textwidth]{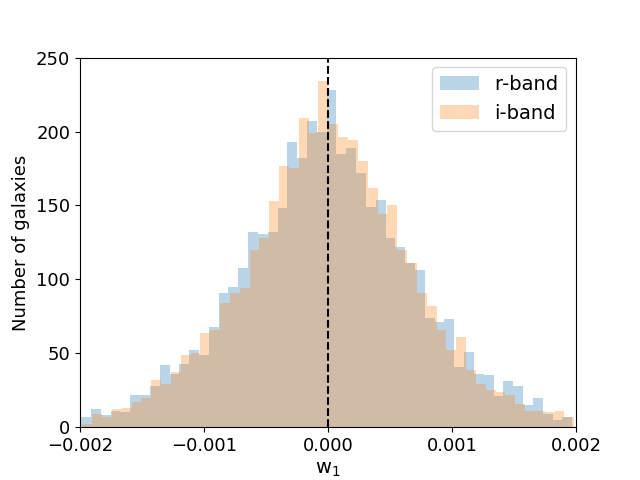}
  \caption{\emph{Left:} Distribution of measured angular gas--star offsets $r_*$ over the 15,634 galaxies in the gas--star offset sample. \emph{Right:} Distribution of measured $r$- and $i$-band warp magnitudes $w_1$ over the 4,149 galaxies in the warp sample.}
  \label{fig:signal_hists}
\end{figure*}

In the left panel of Fig.~\ref{fig:signal_hists} we show the distribution of angular $r_*$ values over our gas--star offset sample, and in the right panel we show the distribution of $w_1$ measured from either the $r$- or $i$-band images. We find a spread in gas--star offsets $\sim$20$^{\prime\prime}$, corresponding to the angular resolution of the Arecibo beam \cite{Haynes_2011}, while measured $w_1$ values are $\mathcal{O}(0.001)$. The warp magnitudes are very similar in the $r$- and $i$-bands; although the $i$-band is less susceptible to attenuation by dust we use the $r$-band for our fiducial analysis to increase consistency with other parts of the model.

\section{Methodology}
\label{sec:method}

Our goal is to derive probability distributions for $\vec{r}_*$ and $w_1$ expected under a screened fifth force of strength $\dG$ and range $\lambda_C$, for each galaxy in the respective samples. By combining this with an astrophysical noise model and comparing with the measured values we will then be able to constrain $\dG$ and $\lambda_C$. Our approach builds on \cite{HIOC, warp}, to which we refer for further detail. We provide an overview of the method here, with particular attention to the parts we have generalised or improved.

\subsection{Modelling the gravitational and fifth-force fields}
\label{sec:gravity_maps}

We begin by calculating the environmental screening potential $\Phi_\text{ext}$ and fifth-force acceleration $\vec{a}_5$ as a function of $\lambda_C$ for each test galaxy. $\Phi_\text{ext}$ is the dimensionless Newtonian potential sourced by masses within $\lambda_C$, which has been shown in simulations to provide a good proxy for environmental screening \cite{Zhao_prl, Zhao_prd, Cabre}. For a set of point masses $j$ this is
\begin{equation}\label{eq:phi}
\Phi_{\text{ext},i}(\lambda_C) = \sum_{r_{ij}<\lambda_C} \frac{G_\text{N} M_j}{r_{ij}},
\end{equation}
at test object $i$. $\vec{a}_5$ is modelled as a Yukawa force from sources within $4\lambda_C$ (as opposed to long-range but cut off at $\lambda_C$, as previously):
\begin{equation}\label{eq:a5}
\vec{a}_{5,i}\left(\dG, \lambda_C\right) = \sum_{r_{ij}<4\lambda_C} -\frac{\Delta G M_j}{r_{ij}^2} \: e^{-r_{ij}/\lambda_C} \: \left(1 + \frac{r_{ij}}{\lambda_C}\right) \: \hat{r}_{ij}.
\end{equation}
These are calculated by modelling the distribution of mass in the local universe using the method of \cite{maps}, as implemented in \cite{HIOC, warp} (for other approaches see \cite{Cabre, Shao}). This procedure has three main steps:

\begin{enumerate}

\item Assign halos to the galaxies observed in a magnitude-limited survey using the technique of halo abundance matching (AM; \cite{Conroy, Kravtsov}), and assume NFW halo profiles to create a first component of the $\Phi_\text{ext}$ and $\vec{a}_5$ fields.

\item Use an N-body simulation to add in statistically the contribution to $\Phi_\text{ext}$ and $\vec{a}_5$ from halos hosting galaxies too faint to be observed by the survey in the previous step.

\item Fill in the mass in long-wavelength modes of the density field associated with very low-mass halos (or not associated with halos at all) using the large-scale inference of the BORG algorithm \cite{BORG_1, BORG_2}.

\end{enumerate}

We now describe each of these steps in turn.

\subsubsection{Primary source catalogue}

Similarly to \cite{HIOC, warp}, we generate 200 mock galaxy catalogues by applying the AM model of \cite{Lehmann} with the S\'{e}rsic $r$-band luminosity function of \cite{Bernardi_SMF} to the \textsc{rockstar} halo catalogue of the \textsc{DarkSky}-400 simulation \cite{DarkSky}. Each catalogue uses different values of the two parameters $\{\alpha, \sigma_\text{AM}\}$ of the AM model drawn from the posterior obtained in \cite{Lehmann} from galaxy clustering, as well as a different realisation of the stochasticity in the galaxy--halo connection implied by given $\sigma_\text{AM}$. For each Monte Carlo realisation of our model, we select a `true' magnitude for each galaxy in the primary source catalogue by scattering the observed magnitude by the uncertainty, and then assign it the halo from a randomly-chosen mock catalogue that is closest to it in magnitude. Repeating the Monte Carlo draws many times thus propagates the uncertainties both in the true magnitudes of the source galaxies and in the galaxy--halo connection into the DM distribution associated with the primary source catalogue, and hence the $\Phi_\text{ext}$ and $\vec{a}_5$ fields that they produce. The screening potential of an NFW halo is
\begin{equation}\label{eq:phi_nfw}
\Phi_\text{ext,NFW}(r) = -\frac{G_\text{N} \mv}{r} \frac{\ln(1 + c r/\rv)}{\ln(1+c) - c/(1+c)},
\end{equation}
while the $\vec{a}_5$ field solves the massive Klein-Gordon equation (Eqs.~\ref{eq:a5_0} \& \ref{eq:mKG}):
\begin{eqnarray}\label{eq:a_nfw}
& & a_{5,\text{NFW}}(r) = -\frac{G_\text{N} \mv}{r^2} \frac{c \: (1+c) \: e^{-(r + \rv/c^2)/\lc}}{{2 \: (\rv + c^2 r) \: ((1 + c) \ln(1 + c) - c)}} \times \nonumber \\
& & \Biggl(e^{2(r + \rv/c^2)/\lc} \left(\frac{r}{\lc} - 1\right) \left(\frac{\rv}{c} + c r\right) \Ei\left(-\frac{r + \rv/c^2}{\lc}\right) \\
& & - \left(1 + \frac{r}{\lc}\right) \left(\frac{\rv}{c} + c r\right) \Ei\left(\frac{r + \rv/c^2}{\lc}\right) \nonumber \\
& & + 2 e^{\rv/(\lc c^2)} \left(c e^{r/\lc} r + \left(1 + \frac{r}{\lc}\right) \left(\frac{\rv}{c} + c r\right) \left(\gamma + \ln\left(\frac{\rv}{\lc c}\right)\right)\right)\Biggr). \nonumber
\end{eqnarray}
$c$ is the halo's concentration, $\Ei$ is the exponential integral function and $\gamma$ is Euler's gamma constant.

Previously we used 2M++ \cite{2mpp} as the primary galaxy catalogue; here we instead use the NSA. This restricts the region of the sky in which our screening maps apply, but since both the warp and gas--star offsets samples explicitly involve NSA information the galaxies are necessarily within the footprint of that catalogue anyway. The advantage is threefold: i) SDSS is much deeper than 2M++, allowing us to explicitly model substructure several orders of magnitude lower in luminosity and hence extend our screening maps out to $\sim$290 Mpc (see below) without crippling statistical uncertainties due to a dearth of observed objects at large distance, ii) the selection in brightness is more homogeneous than 2M++ so that it is easier to define a magnitude limit with near-100\% completeness, and iii) the use of SDSS magnitudes facilitates the use of AM to assign halo masses and concentrations to the source galaxies. This is because the S\'{e}rsic $r$-band luminosity function of SDSS is well-known, and the AM model that we use was specifically calibrated on it.

\subsubsection{Restoring low-mass substructure}

The addition of halos hosting galaxies too faint to be recorded in the NSA is also based on AM mocks. After performing AM on all \textsc{DarkSky} halos with more than 200 particles, we place the observer at the centre of the box and calculate the apparent $r$-band magnitude of each halo. We do this by fitting a linear scaling relation between the K-correction factor $K$ and distance $D$ in the NSA, finding $K = 0.00028366 \: D/\text{Mpc} - 0.00011111$ to provide a good fit. We calculate the apparent magnitude as
\begin{equation}
m_r = M_r + 5 \log(D/\text{Mpc}) + 25 + K
\end{equation}
(where $\log$ denotes $\log_{10}$). We then separate the mock galaxies in the simulation into those that would be visible to the NSA and those that would not, according to whether $m_r < 17.77$. Then for each galaxy with $m_r < 17.77$ we calculate the $\Phi_\text{ext}(\vec{x}; \lambda_C)$ and $\vec{a}_5(\vec{x}; \lambda_C)$ generated both by neighbouring \emph{visible} galaxies alone, $\Phi_\text{ext,vis}$ and $\vec{a}_{5,\text{vis}}$, and by \emph{all} neighbouring halos, $\Phi_\text{ext,all}$ and $\vec{a}_{5,\text{all}}$. The former are the quantities we have access to in the data, while the latter are those we wish to estimate in this step. We again assume NFW profiles and hence use Eqs.~\ref{eq:phi_nfw} and~\ref{eq:a_nfw}.

Previously we used a simple counts-in-cells algorithm to predict the total quantities from the visible ones, which we replace here with a machine learning method capable of describing the complex correlations more faithfully. As well as the quantities we wish to predict, we record for each `visible' \textsc{DarkSky-400} halo a set of features that correlate with these quantities and that we have access to in the data, which we use as predictors. These are the absolute $r$-band magnitude and distance of the galaxy, the number of objects visible to the NSA within $\lambda_C$ and $4\lambda_C$, $\Phi_\text{ext,vis}$ and $|\vec{a}_{5,\text{vis}}|$. In addition to $\Phi_\text{ext,all}$ and $|\vec{a}_{5,\text{all}}|$ we wish to predict the angle through which the fifth-force acceleration is rotated due to mass not visible to the NSA (i.e. between $\vec{a}_{5,\text{vis}}$ and $\vec{a}_{5,\text{all}}$), and the mass $M_\text{invis}$ of these objects within $4\lambda_C$ (which will be required later). For each quantity to be predicted we train a random forest, extratrees or multi-layer perceptron (neural net) regressor with these features, and use the \textsc{GridSearchCV} function of \textsc{scikit-learn} \cite{scikit_learn} to select the algorithm best able to reproduce the training data, varying each of the relevant regression parameters. We find the best fits to be generated by a random forest for all quantities except $\Phi_\text{ext,all}$, which prefers a neural net. We overplot the predictions on the training data to verify explicitly that the optimised trained regressors provide good fits to $\Phi_\text{ext,all}$, $\vec{a}_{5,\text{all}}$ and $M_\text{invis}$ for the vast majority of cases in the training data from the simulation, for all $0.3 < \lambda_C/\text{Mpc} < 8$. We then apply the trained regressors to the real data to predict these quantities from the observable features.

\subsubsection{Modelling the mass outside halos}

Our method for adding mass that would not be associated with halos of at least 200 particles in \textsc{DarkSky} is based on the inference of the BORG-PM algorithm applied to the 2M++ catalogue \cite{Lavaux, BORG-PM}. (There exists a BORG reconstruction using SDSS data \cite{BORG-SDSS}, but this has significantly lower spatial resolution than the 2M++ reconstruction.) The BORG-PM algorithm employs a full particle mesh-based forward model for the $z=0$ DM distribution as a function of galaxy bias parameters and the phases of Gaussian DM initial conditions. By sampling these with a Poissonian likelihood for the number densities of 2M++ galaxies in voxels, constraints are obtained on both the initial conditions of the 2M++ volume and the final density field smoothed on scales $\sim$3.8 Mpc. We take 10 realisations of this density field drawn from the posterior, separated by several autocorrelation lengths. For each Monte Carlo draw from our model we select a random realisation and subtract the masses modelled in the previous steps (down to a minimum voxel mass of 0) to prevent double counting. The masses of halos from step 1 are removed from the grid cells in which they reside, and $M_\text{invis}$, calculated in step 2, is subtracted equally from all grid cells within $4\lambda_C$ of the test point in question. The contribution of the remaining mass is added to our estimates of $\Phi_\text{ext,all}$ and $\vec{a}_{5,\text{all}}$.

Previously we modelled this density field using point masses at the initial grid cell centres; now we interpolate the field onto a finer grid of resolution 1.4 Mpc and then smooth by a Gaussian kernel of width $\sigma_G$ equal to a quarter of the grid cell size. This stabilises the $\Phi_\text{ext}$ and $\vec{a}_5$ fields sourced by this density and eliminates artificial caustics generated near the grid cell centres. The screening potential sourced by a Gaussian is
\begin{equation}
\Phi_\text{ext,Gauss}(r) = -\frac{G_\text{N} M}{r} \: \erf \left(\frac{r}{\sqrt{2} \: \sigma_G}\right),
\end{equation}
while the fifth-force field from the massive Klein-Gordon equation is
\begin{eqnarray}
& & a_{5,\text{Gauss}}(r) = -\frac{G_\text{N} M}{r^2} \: e^{-r/\lc} \: \left(1 + \frac{r}{\lc}\right) \: e^{\sigma_G^2/(2 \lc^2)} \times \\
& & \Biggl(\frac{1}{2} \left(1 + \erf \left(\frac{r-\sigma_G^2/\lc}{\sqrt{2} \sigma_G}\right)\right) - \frac{1}{2} \frac{1-r/\lc}{1 + r/\lc} \left(1 - \erf \left(\frac{r+\sigma_G^2/\lc}{\sqrt{2} \sigma_G}\right)\right) \nonumber \\
& & -\frac{2 r}{\sqrt{2 \pi} \: \sigma_G \: (1+r/\lc)} \: e^{-(r-\sigma_G^2/\lc)^2/(2 \sigma_G^2)}\Biggr). \nonumber
\end{eqnarray}

\subsubsection{Final considerations}

The final contribution to galaxies' screening potentials, which we denote $\Phi_\text{in}$, derives from their own masses. We now use a homogeneous formula for this across both test and source galaxies, which has been shown in simulations to be a reliable predictor for self-screening \cite{Zhao_prl, Cabre}:
\begin{equation}
\Phi_\text{in} = -V_\text{vir}^2.
\end{equation}
$V_\text{vir}$ is the virial velocity of the halo, determined by the AM association between the galaxy in question and a halo in the \textsc{DarkSky-400} simulation. We reiterate that an object is considered screened if $|\Phi_\text{tot}| \equiv |\Phi_\text{ext}| + |\Phi_\text{in}| > \chi$ (Sec.~\ref{sec:f(R)}), and that in $f(R)$ gravity $\chi = 3/2 \: f_{R0}$.

Only unscreened masses source $\vec{a}_5$. We determine which halos in step 1 are unscreened by evaluating $\Phi_\text{ext}(\vec{x}; \lambda_C)$ at their positions and adding their $\Phi_\text{in}$. To determine which portions of the smoothed density field in step 3 are unscreened, we interpolate the unscreened fraction over the source halos using a multidimensional piecewise linear interpolator, and evaluate this fraction at the positions of the smoothed density field grid cell centres. The NSA source halos populate the test volume sufficiently densely for this interpolation of $\Phi_\text{ext}$ not to introduce significant uncertainty.

The maximum extent of our screening maps is 290 Mpc because at larger distance the 2M++ survey is substantially incomplete and hence the density field reconstructed by BORG becomes unreliable \cite{2mpp, Lavaux}. To ensure that the fifth-force field at the position of the most distance test point that we consider is robust, we need to model sources up to $4\lambda_{C,\text{max}}$ more distant than it. However, whether or not the objects sourcing this field are screened depends on mass a further $\lambda_C$ distant, with the result that the screening map must extend $5\lambda_{C,\text{max}}$ further than the test galaxy sample. We restrict our attention here to the lower end of the $\lambda_C$ range considered in \cite{HIOC, warp}, both because it corresponds to weaker modified gravity (e.g. lower $f_{R0}$, less stringently constrained by past studies) and because this increases the distance range over which we can include test galaxies. In particular, choosing $\lambda_{C,\text{max}}=8$ Mpc means that with a screening map extending to 290 Mpc we can consider data only out to 250 Mpc. This explains the distance cut used in Sec.~\ref{sec:data}.

\subsection{Modelling the halo restoring force}
\label{sec:model_signals}

With $\Phi_\text{tot}(\lambda_C)$ and $\vec{a}_5(\lambda_C)$ for each galaxy in hand, the final piece of information required to calculate the fifth-force expectation for $\vec{r}_*$ (Eq.~\ref{eq:rstar}) and $w_1$ (Eq.~\ref{eq:w1_integral}) is the restoring force due to the test galaxies' halos. In \cite{HIOC}, we assumed for the gas--star offset analysis a uniform DM density within $r_*$ with magnitude determined by an empirical scaling relation with baryonic surface mass density. This has the disadvantage first of not calibrating the DM density against a simulation, potentially introducing a discrepancy with other parts of the model, and second of introducing a significant degeneracy between the fifth-force constraint and the unknown scatter in the central density (see \cite{HIOC} fig. 16). We therefore replace that model here with a power-law density (as used in \cite{warp}):
\begin{equation}\label{eq:density}
\rho(r) = \rho_{rs} \: (r/r_s)^{-\beta}
\end{equation}
with $\rho_{rs} \equiv \rho(r_s)$ and $\beta$ a variable exponent. This gives an enclosed mass
\begin{equation}\label{eq:M_enc}
M(r) = \frac{4 \pi \: \rho_{rs}}{3-\beta} \: r_s^\beta \: r^{3-\beta}.
\end{equation}
Plugging into Eq.~\ref{eq:rstar} we find the expression for $r_*$ under a fifth force:
\begin{equation}\label{eq:rstar_2}
\vec{r}_* = \left(a_5 \: \frac{\Delta G}{G_\text{N}^2} \: \frac{3-\beta}{4 \pi \rho_{rs}} \: r_s^{-\beta}\right)^\frac{1}{1-\beta} \hat{a}_5.
\end{equation}
Plugging instead into Eqs.~\ref{eq:w1_curve} and~\ref{eq:w1_integral} gives the expectation for $w_1$:
\begin{equation}\label{eq:w1_final}
w_1 = - \frac{\beta \: (3-\beta)}{(\beta+1)(\beta+2)} \: a_{5,z} \: \frac{\Delta G}{G_\text{N}^2} \: \frac{1}{4 \pi \rho_{rs}} \: \frac{(3 R_\text{eff}/r_s)^\beta}{3R_\text{eff}}.
\end{equation}

We calculate $\rho_{rs}$ for each test galaxy as the density at the scale radius of the NFW halo associated to the galaxy by AM (after first subtracting the stellar mass). In each Monte Carlo realisation of the model the galaxy is associated to a different halo and hence the uncertainties on $r_s$ and $\rho_{rs}$ are naturally propagated. $\beta\simeq1$ corresponds to an NFW profile (at $r<r_s$ where galaxies are situated), but this may be altered by the process of galaxy formation. In particular the adiabatic contraction of DM towards the halo centre as baryons collapse to form the galaxy will increase $\beta$ \cite{Blumenthal, Gnedin}, but subsequent stellar feedback that injects energy into the halo may reduce it \cite{Pontzen_Governato, DC, DP_CuspCore}. Observational evidence has been reported for a wide range of $\beta$ values (e.g.~\cite{Cusp_core, Sonnenfeld, Oman}) although there is a tendency for values $<1$ to be preferred, especially at low masses (the `cusp-core problem' \cite{Cusp_core}). We therefore take $\beta=0.5$ as our fiducial case, although we check in Sec.~\ref{sec:discussion} that any reasonable value in the range $0 \lesssim \beta \lesssim 1.5$ produces results differing by at most a factor of 2 (although note that $\beta=1$ is singular in Eq.~\ref{eq:rstar_2}). See Sec.~\ref{sec:discussion} and secs. II and IV of \cite{HIOC, warp} for further discussion of the approximations used in this section.

\subsection{Likelihood and noise model}
\label{sec:likelihood}

We are now in a position to calculate the expected $\vec{r}_{*}$ and $w_1$ for each galaxy in the corresponding samples as a function of the fifth-force parameters. As the predicted signals are simple functions of $\dG$ from Eqs.~\ref{eq:rstar_2} and~\ref{eq:w1_final}, the cleanest way to perform the inference is to calculate a `template' likelihood for each galaxy and $\lambda_C$ value at $\dG=1$, and then scale the result according to the particular $\dG$ value one is testing. By repeating the $\dG=1$ calculation of $\vec{r}_{*}$ and $w_1$ 1000 times, in each case randomly sampling from the probability distributions describing the input quantities, we build up discrete empirical representations of the galaxy-by-galaxy likelihood functions for the observables at any $\lc$ of interest.

\begin{figure*}
  \centering
  \includegraphics[width=0.49\textwidth]{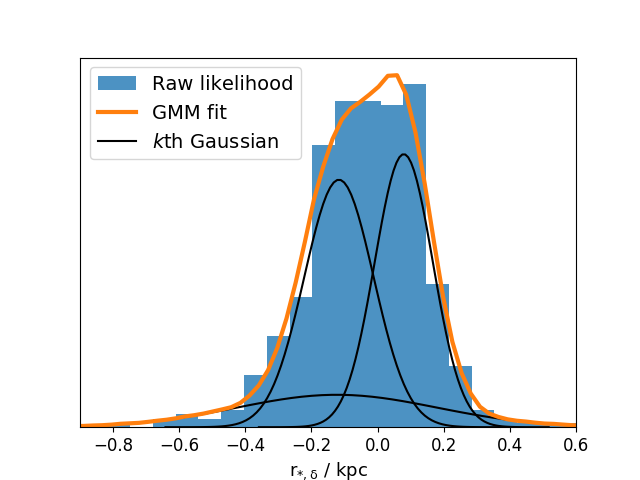}
  \includegraphics[width=0.49\textwidth]{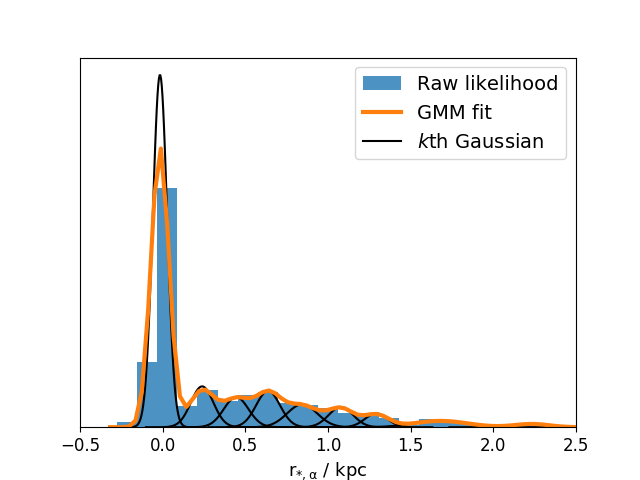}
  \caption{Example Monte Carlo likelihoods with Gaussian mixture model fits overplotted. \emph{Left} shows a typical galaxy with 3 Gaussian components while \emph{right} shows a more extreme one with 11; both come from the gas--star offset sample with $\dG=1$ and $\lambda_C=2.4$ Mpc.}
  \label{fig:GMM_fit}
\end{figure*}

This must then be used to estimate a continuous likelihood function in order to apply Bayes' theorem to derive posteriors on the model parameters. Previously we used a 10-bin histogram of the Monte Carlo data for this purpose, which has the disadvantage first of fairly low resolution and second of potential sensitivity to outliers at the histogram edges. We therefore replace this here with a Gaussian mixture model (GMM). Specifically, we use the \textsc{scikit-learn} function \textsc{GaussianMixture} to fit $N_\text{G}$ 1D Gaussians with free weights, means and variances to the 1000 $r_{*,\alpha}$, $r_{*,\delta}$ and $w_1$ values for a given galaxy. This is done for $\dG=1$ and separately for 21 $\lambda_C$ values in the range $0.3-8$ Mpc. We calculate the Bayesian Information Criterion (BIC) of each fit and select the value of $N_\text{G}$ that minimises this, up to a maximum of 25. We find typical $N_\text{G}$ values of $2-6$ with only a small tail to larger $N_\text{G}$. The fitted GMMs provide a good description of the Monte Carlo data, as illustrated in Fig.~\ref{fig:GMM_fit}. We have also checked that replacing the GMM with a Kernel Density Estimator (KDE) gives similar results.

The fifth-force part of the likelihood for galaxy $i$ is thus given by
\begin{align} \label{eq:L_F5}
&\mathcal{L}_{5,i}(y_i|\dG, \lambda_C) = (1-f_i) \: \delta(y_i) \\ \nonumber
&+ f_i \: \sum_{k=1}^{N_\text{G}} \: \frac{W_{ik}}{\sqrt{2\pi} \: F[\dG] \: s_{ik}} \: \exp\left\{-\frac{(F[\dG] \: \bar{y}_{ik} - y_i)^2}{2 F[\dG]^2 \: s_{ik}^2}\right\},
\end{align}
where $y \in \{r_{*,\alpha}, r_{*,\delta}, w_1\}$, $f_i$ is the galaxy's unscreened fraction (fraction of Monte Carlo model realisations in which $|\Phi_\text{tot}|<\chi$), and $W_{ik}$, $\bar{y}_{ik}$ and $s_{ik}$ are the fractional weight, mean and width of Gaussian $k$ in the mixture model. $F[\dG]$ describes the scaling of the template signal: $F[\dG] \equiv (\dG)^{1/(1-\beta)}$ for the gas--star offset inference (Eq.~\ref{eq:rstar_2}) and $F[\dG] \equiv \dG$ for the warp inference (Eq.~\ref{eq:w1_final}).

A fifth force is however not the only phenomenon relevant to our modelling: a variety of `baryonic physics' effects including ram pressure, stellar feedback, mergers, tidal effects and secular evolution could lead to non-zero measured $\vec{r}_*$ and $w_1$ values, in addition to observational uncertainties associated with the resolution of the observations. To model these effects we convolve the fifth-force part of the likelihood function (i.e. each of the Gaussians in the GMM) with an additional Gaussian of width $\sigma_i$ in physical space for galaxy $i$. There are three basic possibilities for setting $\sigma_i$, which may be combined:
\begin{enumerate}
\item $\sigma_i = \sigma_\text{const}$ is the same for all galaxies, representing a universal Gaussian noise distribution for the signal of which each galaxy is a realisation,
\item $\sigma_i = \sigma_D \, D_i$ is proportional to the galaxy's distance, describing an angular resolution for the measurement,
\item $\sigma_i$ is a general function of multiple galaxy parameters, describing an astrophysical contribution to the signals that depends on the properties of the galaxies and/or their environments.
\end{enumerate}

We showed in \cite{HIOC} that including a dependence of $\sigma_i$ on the signal-to-noise ratio of the ALFALFA detection led to no significant impact on the $\dG$ posterior, because the information on $\dG$ derives from galaxy-by-galaxy correlations of the measured $\vec{r}_*$ with that predicted, rather than from the overall shape of the measured $\vec{r}_*$ distribution as modelled by the noise terms. In \cite{warp} we allowed for variations of $\sigma_i$ with the minor-to-major axis ratio $b/a$ of the disk, which we found to correlate with the measured $w_1$ in a way that was difficult to model. This motivated our present choice to focus exclusively on galaxies with $b/a=0.15$, obviating this potential systematic. In Sec.~\ref{sec:systematics} we provide further evidence that baryonic effects are not a significant systematic in our inference, leading us not to pursue more complex noise models of the $3^\text{rd}$ class.

A combination of options 1 and 2 is a priori plausible for both the warp and gas--star offset signals. To determine which terms to include, we compare the maximum-likelihood values of the inferences with distance-independent noise, distance-dependent noise, and both. Specifically, we calculate the BICs of the three models for several $\lambda_C$ values spanning our range of interest, as a measure of their goodness of fit relative to the number of free parameters they contain. For the gas--star offset signal we find that a distance-dependent noise (option 2) performs far better than distance-independent noise (option 1) in terms of maximum likelihood, and combining them gives no improvement over the former. This accords with our intuition that the magnitude of the measured gas--star displacement is determined primarily by the angular resolution of the Arecibo beam. For the warp signal we find the distance-independent noise to perform moderately better than the distance-dependent noise, and the improvement on combining them (with a best-fit distance-dependent noise term $>$0 at $\sim$3$\sigma$) to be marginal. For simplicity we therefore opt for only a distance-independent noise term in the warp inference, although we have verified explicitly that the $\dG$ constraints change by at most 50\% if both terms are used. We conclude that galaxies' apparent warps are set predominantly by stochastic intrinsic and/or environmental galactic processes, with little dependence on the spatial resolution of the instrument that observes them.

We treat $\sigma_{r_*} \equiv \sigma_D$ for gas--star offsets and $\sigma_{w_1} \equiv \sigma_\text{const}$ for warps as free parameters in the respective inferences, marginalising over them in deriving constraints on the fifth-force parameters. This essentially broadens the Gaussians in the GMM into student-t distributions with heavier tails, weakening the $\dG$ constraints and making the analysis conservative. The final noise models provide good fits to the measured distributions, with negligible improvement on adding further terms from e.g. an Edgeworth expansion~\cite{Edgeworth}. The full likelihood is given by
\begin{align} \label{eq:L_tot}
&\mathcal{L}_i(y_i|\dG, \lambda_C, \sigma_i) = (1-f_i) \: \frac{\exp\{-y_i^2/2\sigma_i^2\}}{\sqrt{2 \pi} \: \sigma_i} + f_i \sum_{k=1}^{N_\text{G}}\\ \nonumber
&\frac{W_{ik}}{\sqrt{2 \pi (F[\dG]^2 \: s_{ik}^2 + \sigma_i^2)}} \exp\left\{-\frac{(F[\dG] \: \bar{y}_{ik} - y_i)^2}{2(F[\dG]^2 \: s_{ik}^2 + \sigma_i^2)}\right\},
\end{align}
i.e. Eq.~\ref{eq:L_F5} with the delta-function for screened galaxies replaced by a Gaussian of width $\sigma_i$ and $F[\dG] s_{ik} \rightarrow \sqrt{F[\dG]^2 s_{ik}^2 + \sigma_i^2}$.

We are now ready to derive the posteriors on $\{\dG, \sigma_{r_*}, \sigma_{w_1}\}$ at each $\lambda_C$ using the \textsc{emcee} sampler \cite{emcee}. We treat $r_{*,\alpha}$ and $r_{*,\delta}$ and different galaxies as independent, multiplying their likelihoods. We impose the priors $\dG, \sigma_{r_*}, \sigma_{w_1} \ge 0$ and sample with $\sim$30 walkers, terminating when the chain is at least 50 autocorrelation lengths long in all parameters and the estimates of the autocorrelations lengths change by $<2\%$ per iteration. We consider the warp and gas--star inferences separately and also combine their constraints by multiplying the total likelihoods.

\vspace{4mm}

\begin{figure*}
  \centering
  \includegraphics[width=0.49\textwidth]{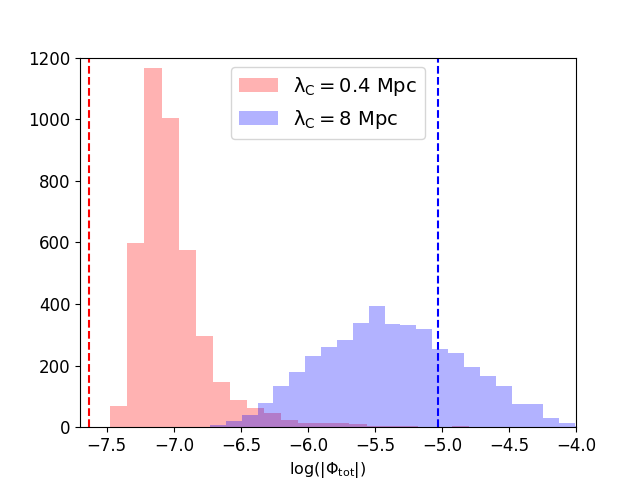}
  \includegraphics[width=0.49\textwidth]{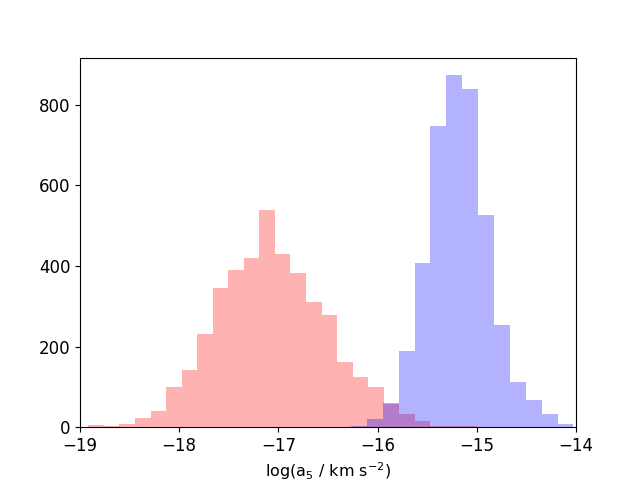}
  \includegraphics[width=0.49\textwidth]{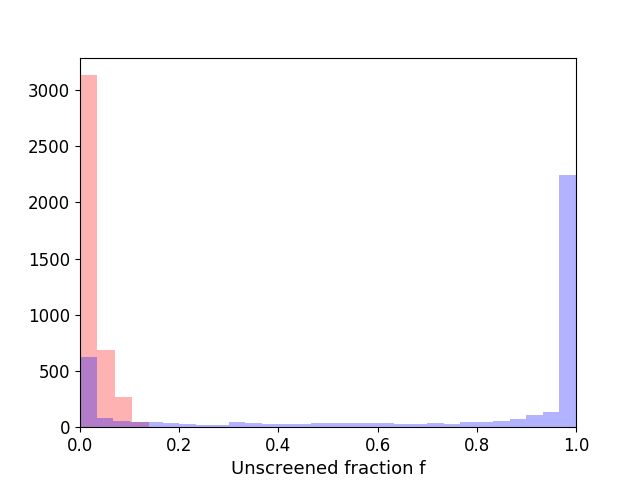}
  \includegraphics[width=0.49\textwidth]{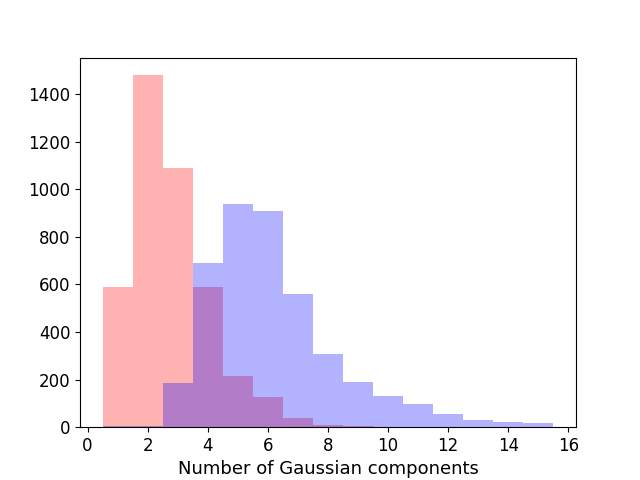}
  \caption{Histograms of various quantities in the warp sample at $\lambda_C=0.4$ Mpc (red) and $\lambda_C=8$ Mpc (blue). \emph{Upper left:} Modal total screening potential (environmental + internal) over the Monte Carlo model realisations. The dashed vertical lines indicate the threshold values for screening in the two models ($\chi$; Eq.~\ref{eq:chi_lambdaC}). Note that although on average $|\Phi_\text{tot}|$ always exceeds $\chi$ for $\lc=0.4$ Mpc this is not necessarily the case for individual realisations, which can lead to non-zero unscreened fractions. \emph{Upper right:} Magnitude of fifth-force field. \emph{Lower left:} Fraction of Monte Carlo realisations in which $|\Phi_\text{tot}|<\chi$ for a given galaxy. \emph{Lower right:} Number of Gaussian mixture model components fitted to the $w_1$ likelihood (Eq.~\ref{eq:L_F5}).}
  \label{fig:signals}
\end{figure*}

\noindent In Fig.~\ref{fig:signals} we show the distributions of screening potential, fifth-force strength, galaxy unscreened fraction and number of Gaussian likelihood components produced by our model over the galaxies in the warp sample, separately for $\lambda_C=0.4$ Mpc and $\lambda_C=8$ Mpc. A greater range for the fifth force implies a greater gravitational potential and fifth-force field as these are sourced by more masses, and the greater self-screening parameter means that galaxies tend to be more unscreened. The fact that at $\lambda_C=8$ Mpc the majority of galaxies in the sample are unscreened means that raising $\lambda_C$ further would only weakly increase the number of galaxies providing constraining power and hence the strength of the constraints, while for $\lc < 0.3$ Mpc practically all objects are fully screened so no constraints are possible. The empirical likelihood becomes slightly more complex as larger volumes contribute to a galaxy's fifth-force properties, resulting in an increase in the best-fit number of Gaussian components. The distributions interpolate between those illustrated for $\lc$ between 0.4 and 8 Mpc, and are qualitatively similar for the gas--star offset inference.

\section{Results}
\label{sec:results}

Our key results are given in Fig.~\ref{fig:constraints}, where we show the $1\sigma$ upper limits on $\dG$ as a function of $\lambda_C$ (or alternatively $f_{R0}$ as shown on the top axis) separately for the gas--star offset, warp and joint analyses. Interpolating the joint constraint to $\dG=1/3$ we find a $1\sigma$ constraint on $f_{R0}$ of $1.4 \times 10^{-8}$, and a similar exercise with the $2\sigma$ line gives $f_{R0} < 2.8 \times 10^{-8}$. The rest of this section describes and elaborates on our results, while the following section discusses potential systematics and places our findings in a broader context.

Fig.~\ref{fig:constraints} shows that both signals contribute to our final constraint, with the warp signal more constraining at $\lambda_C \lesssim 2.5$ Mpc and gas--star offsets at $\lambda_C \gtrsim 2.5$ Mpc. This is caused by a trade-off between the greater number of galaxies in the gas--star offset sample, the distance-independence of the noise term in the warp inference, and the relative magnitudes of the predicted vs observed signals in the two analyses. This implies that a multi-tracer approach to constraining fundamental physics with galaxies is beneficial, and we anticipate even greater gains with future data at a range of wavelengths.

The degree of bumpiness of the constraint lines indicates the overall noise level of the experiment -- and hence uncertainty on our constraints -- due to the finite number of Monte Carlo realisations used to reconstruct the likelihood. To quantify this we repeat the end-to-end warp inference 8 times at $\lambda_C=0.4$, finding an unbiased sample variance in the $1\sigma$ and $2\sigma$ $\dG$ constraints of 0.014. This is a 5.8\% uncertainty on the mean $1\sigma$ constraint of 0.241, showing that the number of Monte Carlo model realisations (1000) is sufficient for the results to be well-determined. In particular this $\lc$ value ($f_{R0}=1.6 \times 10^{-8}$) remains fully ruled out at $1\sigma$.

\begin{figure}
   \centering
   \includegraphics[width=0.99\linewidth]{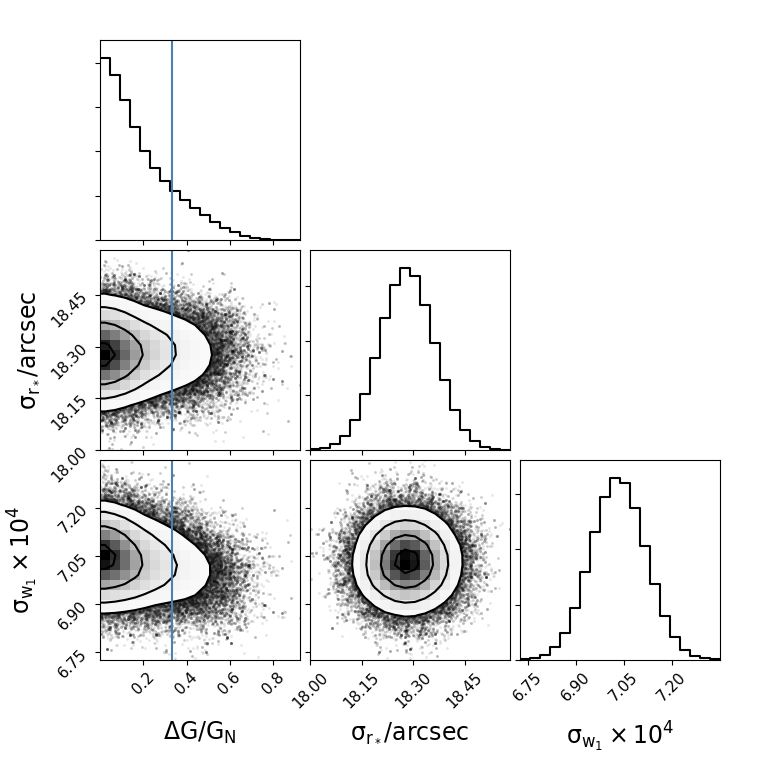}
   \caption{Corner plot of the constraints on the model parameters ($\dG$ and the two noise terms) for the joint analysis of gas--star offsets and warps at $\lambda_C=0.4$ Mpc, corresponding to $f_{R0}=1.6\times10^{-8}$. The blue vertical line indicates $\dG=1/3$, which is ruled out at $>1\sigma$ confidence.}
   \label{fig:corner}
\end{figure}

In Fig.~\ref{fig:corner} we show the corner plot from the joint inference at $\lambda_C=0.4$ Mpc. We find, as in \cite{HIOC, warp}, that the noise terms $\sigma_{r_*}$ and $\sigma_{w_1}$ have very little degeneracy with $\dG$. This is because they describe the overall shapes of the $\vec{r}_*$ and $w_1$ distributions, while the $\dG$ constraint is set by the correlations (if any) between the measured signals and those predicted. We recover the $\sim$18$^{\prime\prime}$ angular uncertainty in HI centroid due to the HI beam width (\cite{Haynes_2011} \& Fig.~\ref{fig:signal_hists}), and find a standard deviation of measured $w_1$ values around 0.0007. The corner plots are qualitatively similar for all $\lambda_C$ values (and for the warp and gas--star offset analyses separately), with only the scale on the $\dG$ axis changing.

\begin{figure}
  \centering
  \includegraphics[width=0.49\textwidth]{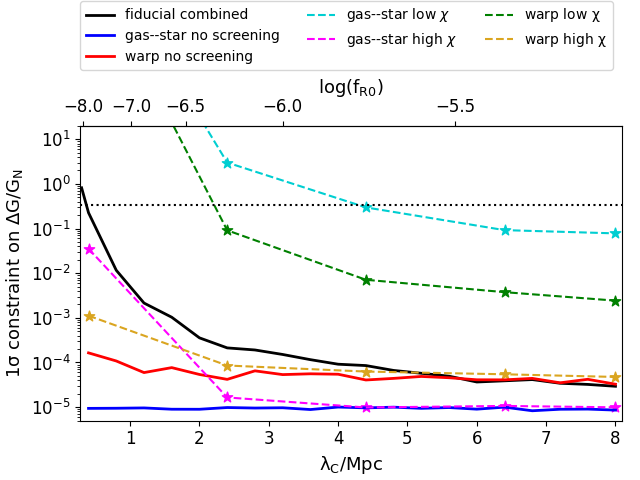}
  \caption{As Fig.~\ref{fig:constraints}, but for a different set of screening scenarios. The blue and red lines show constraints from the gas--star offset and warp analyses with screening switched off, so that all test galaxies are considered unscreened and all source masses contribute to $\vec{a}_5$. The cyan, magenta, green and gold lines show constraints with $\chi$ raised or lowered by a factor of 10 from its fiducial dependence on $\lambda_C$ (Eq.~\ref{eq:chi_lambdaC}). In this case the $\dG$ limits are evaluated at $\lambda_C=0.4,2.4,4.4,6.4$ and 8 Mpc (marked by stars) and connected by piecewise linear interpolation. The black solid line is reproduced from Fig.~\ref{fig:constraints}.}
  \label{fig:constraints_2}
\end{figure}

In Fig.~\ref{fig:constraints_2} we show in the red and blue solid lines the results with screening turned off, so that all source masses contribute to $\vec{a}_5$ and the unscreened fraction of test galaxies is pinned to 1 in Eqs.~\ref{eq:L_F5} and~\ref{eq:L_tot}. The $\dG$ constraints are significantly stronger in this case because $a_5$ is increased and all test galaxies contribute constraining power, and the dependence on $\lambda_C$ is reduced because this no longer determines galaxies' degrees of screening through $\chi$. The gas--star offset constraints are stronger than the warp constraints for all $\lambda_C$ without screening due to the larger sample size, so that the joint constraints coincide with them.

It is important to bear in mind however that this constraint requires the fifth force to couple differently to stars, gas and dark matter, which in the absence of screening is not naturally the case. Note also that without screening the warp and gas--star offset results apply to slightly different models: in the former the dark matter must couple to the fifth force but not the stars, or vice versa (and the gas is practically irrelevant), while in the latter the stars and gas must couple differently while it does not matter which one the dark matter follows. An example of the former is a model in which the fifth-force operates only in the dark sector; we conclusively rule out, for example, the $\dG=\mathcal{O}(0.1)$ proposed in \cite{Vafa}. These constraints apply in general to coupled quintessence models \cite{Wetterich_cosmon, Amendola, Copeland}.

Next we investigate the effect of varying the relation between $\lambda_C$ and the self-screening parameter $\chi$. Although related one-to-one in Hu-Sawicki $f(R)$ (Eq.~\ref{eq:chi_lambdaC} for $n=1$) these are independent variables in general thin-shell models, with $\lambda_C$ controlling the strength and orientation of the fifth-force field over space and $\chi$ controlling objects' degrees of screening. General chameleon models obey the relation \cite{Compendium}
\begin{equation}\label{eq:chi_lambdaC_general}
\chi = \frac{1}{2} \: (1+p) \: \lc^2 \: \rho,
\end{equation}
where $p$ is the power-law index of the scalar field in the potential: $V(\phi) \sim \phi^{-p}$. The Hu-Sawicki model with $n=1$ corresponds to $p=-1/2$, while another popular choice, $n=3$, corresponds to $p=-3/4$. Eq.~\ref{eq:chi_lambdaC_general} then shows that the $\chi$ values for these models are normalised differently by a factor of 2, with $n=3$ giving lower $\chi$ and hence making objects easier to screen. Beyond $f(R)$ another commonly-studied case is $p=1$, giving a larger $\chi$ than $n=1$ Hu-Sawicki by a factor 4.

To understand the effects of these differences on our constraints, we show in the dashed lines of Fig.~\ref{fig:constraints_2} the results when $\chi$ is varied by a factor of 10 on either side of its fiducial value from Eq.~\ref{eq:chi_lambdaC}, bracketing plausible variations in the $\chi-\lc$ relation. A lower $\chi$ means more galaxies screened, reducing $a_5$ and the number of test galaxies contributing constraining power and hence weakening the bound on $\dG$. When $\chi$ becomes very small practically all objects become screened, causing the green and cyan lines to blow up at small $\lambda_C$. Conversely, when $\chi$ is greatly raised the bounds tend to their no-screening values as all objects become unscreened. Even with $\chi$ reduced by a factor of 10 we require $\lc \lesssim 2$ Mpc at $\dG=1/3$, and with $\chi$ reduced by a factor of only 2 ($n=3$ Hu-Sawicki) our constraint on $f_{R0}$ remains better than $10^{-7}$. This again is sufficient to ensure that the great majority of astrophysical objects are screened. Any thin-shell model that can be mapped to the triplet $\{\dG, \lambda_C, \chi\}$ may be similarly constrained. This includes the symmetron and environmentally-dependent dilaton, but as these screening mechanisms operate in a substantially different way to the chameleon the precise bounds on their parameters set by galaxy morphology must be worked out separately.

Finally, we validate our inference framework by generating mock data with a fifth-force signal injected into the best-fit astrophysical noise and checking that our pipeline could detect $\dG$ values as low as we claim to exclude. We investigate two cases: a signal at $\dG=1/3$, $\lc=0.8$ Mpc ($f_{R0}=6 \times 10^{-8}$) in the warp data, or one at $\dG=0.001$, $\lambda_C$ = 8 Mpc in the gas--star offset data. Replacing the real data with these synthetic samples we reconstruct $\dG=0.28 \pm 0.03$ in the first case and $\dG=0.001 \pm 0.0001$ in the second, showing that if either of these scenarios were realised we would have discovered them at $\sim$10$\sigma$ confidence.

\section{Discussion}
\label{sec:discussion}

\subsection{Systematic uncertainties}
\label{sec:systematics}

A primary potential systematic is the model for the restoring force on the stellar disk due to its offset from the halo centre (Sec.~\ref{sec:model_signals}). The relevant DM density is on $\lesssim$ kpc scales, where it is quite uncertain. We have set the halo scale radius and density at that radius using AM (including propagation of its statistical uncertainties), and introduced an additional degree of freedom $\beta$ to describe potential deviations from the NFW $\rho_\text{DM} \propto 1/r$ at $r \ll r_s$ (Eq.~\ref{eq:density}). Our fiducial $\beta$ is 0.5, but to check sensitivity to it we have repeated the inferences with $\beta=0,1,1.5$, finding at most a factor 2 effect on the $\dG$ constraints. It is possible that as well as flattening the density profile within $r_s$, stellar feedback could lower $\rho_{rs}$ itself by pushing dark matter even further out. This would strengthen the $\dG$ constraints since the predicted signals scale with $(\dG)/\rho_{rs}$ (Eqs.~\ref{eq:rstar_2}-\ref{eq:w1_final}), making our analysis conservative.

Under thin-shell screening we expect the gas to follow the DM as it becomes displaced from the stars, so that the restoring force on the stellar component receives contribution from both the gas and DM mass. Our method implicitly assumes that the gas density profile is similar to the DM's as we do not model these two components separately. Galaxies' baryonic masses rarely extend beyond the halo scale radius, so to first order the effect of the galaxy mass itself is captured by an increase of $\beta$. (In any case gas typically provides a subdominant mass contribution relative to DM.) Nevertheless there is no convincing a priori reason to expect power-law density profiles within galaxies -- or that the pivot scale $r_s$ is as predicted by N-body simulations -- so in the future it may be worth repeating the inferences with more detailed galaxy-by-galaxy empirical mass models.

A second key potential systematic is the noise model. Although we have now explored more thoroughly which types of noise term are preferred by the data, we continue to assume that astrophysical contributions to the signals do not correlate significantly with any galaxy property. We have investigated this in two ways.  First, we have optimised and trained a random forest regressor on our datasets to calculate the feature importances of a number of galaxy properties for the prediction of the warp and gas--star offset signals. These features are the distance to the galaxy, $r$-band absolute magnitude, effective S\'{e}rsic radius, HI linewidth, HI mass and signal-to-noise ratio of the ALFALFA detection. We find that $\vec{r}_*$ (when expressed as a physical displacement) correlates significantly only with the galaxy distance, while no feature is particularly useful in predicting $w_1$. This shows that there are not likely to be important components missing from our empirical noise model.

We have also performed a similar analysis on mock galaxies from cosmological hydrodynamical simulations. These allow us to estimate the distributions of the fifth-force signals -- and their correlations with other galaxy properties -- given complete prescriptions for the behaviour of baryons. We calculated $\vec{r}_*$ and $w_1$ in a sample of well-resolved galaxies from the $z=0$ snapshot of the Horizon-AGN simulation \cite{H-AGN}, in addition to their stellar and gas masses and sizes, and virial masses. Beyond a moderate correlation of $r_*$ with effective radius we again find these features to have little importance in the prediction of $\vec{r}_*$ and $w_1$. In particular the distribution of $r_*$ in the simulation is to first order fully described by the spatial resolution of the simulation, $1-2$ kpc. We find similar results in the 75 Mpc/h box of the Illustris-TNG simulation \cite{TNG}. The neglect of `galaxy formation physics' is therefore unlikely to significantly bias our inference. A full investigation of fifth-force signals in hydrodynamical simulations will be presented shortly (Bartlett et al 2020, in prep).

Finally, although our analysis assumes a $\Lambda$CDM cosmology, $f_{R0} \approx 10^{-8}$ has only minute effect on the cosmic expansion history and growth of structure \cite{Lombriser_rev, Arnold} (indeed, this is what makes astrophysical tests of screened fifth forces far more powerful than cosmological ones in general). This assumption is therefore not a potential source of error. Other systematic uncertainties in common with our previous analyses are described in sec. VIB of \cite{HIOC, warp}, which we show there not to be significant.

\subsection{Comparison with the literature}
\label{sec:literature}

Although we find no preference here for $\dG>0$ at any $0.3 < \lambda_C/\text{Mpc} < 8$, we confirm the result of \cite{HIOC, warp} that -- with the selection criteria of those analyses -- there is a preference relative to GR for a screened fifth force with $\lambda_C \simeq 2$ Mpc and $\dG \simeq 0.02$. The difference stems from the galaxies we admit into the respective samples. In \cite{HIOC} we required at most a $2'$ angular separation between the centroids of HI and optical emission, and in \cite{warp} we imposed no maximum to the observed $w_1$ we would allow. We find that with these relatively loose selection criteria, there exist $\lesssim10$ galaxies with extreme measured signals that dominated our previous inferences. The $\Delta \log(\mathcal{L})$ values for these galaxies between $\dG=0.02$ and 0 were $\gtrsim2$ orders of magnitude larger than the rest of the sample, and removing them efficiently eliminates the peaks in \cite{HIOC} fig. 8 and \cite{warp} fig. 11.

These findings motivate the more stringent data quality cuts that we use here: an HI--OC separation $<1^\prime$, and a measured $|w_1|$ value $< 0.002$. These cuts remove 474 and 196 galaxies respectively (2.9\% and 4.5\% of the total samples), and correspond to increased purity with respect to reliable HI--optical association and greater regularity in the disk images. All galaxies now  contribute $\Delta \log(\mathcal{L})$ values of the same order of magnitude, so that the inference is driven by the full galaxy population. As further evidence in favour of these more stringent selection criteria, we note that the resulting constraint excludes $\{\lambda_C \simeq 2 \: \text{Mpc}, \dG \simeq 0.02\}$ at $\sim$3$\sigma$. This suggests that the excluded galaxies are genuine outliers, since the remainder of the population cannot support the scenario they prefer. In any case, given the significant and nontrivial astrophysical uncertainties we do not believe it is possible to infer a fifth force robustly from only a handful of galaxies, but rather this must be done statistically over $\gtrsim\mathcal{O}(10^2)$ objects so that the noise model can be reliably constrained and marginalised over. Nevertheless it would be useful to understand further the peculiar properties of the galaxies that prefer a fifth force, possibly with an eye to more detailed follow-up observations.

Our constraints on $\dG$ are almost an order of magnitude stronger than those of \cite{HIOC, letter} for most $\lambda_C$ due to the larger sample size, more realistic noise model and combination of warp and gas--star offset warp data. Those constraints themselves were a factor of several stronger than any presented previously \cite{Jain_cepheids, Vikram, Lombriser_rev}. For $\chi \lesssim 10^{-8}$, as we require here, even dwarf galaxies in voids become screened, as do the outer layers of the most diffuse stars. Indeed, there are \emph{no} astrophysical objects with screening potentials much below $10^{-8}$, so for all intents and purposes the \emph{entire universe} is screened and modified gravity of this type is of absolutely no astrophysical consequence. While couched in terms of $f(R)$ our results apply to \emph{all} thin-shell screening models and thus provide an experimental standard for every theory in that class. Other theories besides Hu-Sawicki $f(R)$ may also be ruled out.

Our findings alter the outlook of future experiments aiming to constrain modified gravity. Current forecasts for Stage IV large-scale structure surveys indicate that they will reach sensitivity to  $f_{R0} \sim 10^{-6}$, and only with unrealistically accurate models for non-linear clustering (including baryonic feedback) will they be able to probe the smaller scales necessary to do better (e.g. \cite{Pratten, Cautun, Euclid_Forecast}). Our constraints therefore imply that such experiments will not find $f(R)$-type modified gravity, allowing effort to be channelled along other avenues. The dearth of astrophysical objects with $|\Phi_\text{tot}| < 10^{-8}$ means that future bounds on thin-shell screening -- with the sole aim of probing generic extensions to GR -- will have to come from laboratory tests \cite{Lab_F5, Burrage_Sakstein}.

In addition to screened theories we provide strong constraints on any fifth force that has an intrinsically different coupling to different components of galaxies. In particular our warp analysis requires $\dG < 3 \times 10^{-5}$ in the case of a relatively long-range fifth force operating only in the dark sector, which applies to models of coupled quintessence. This bound is comparable in strength to laboratory and Solar System tests of gravity \cite{Adelberger,Cassini,Will}, although we stress that we are not sensitive to the strength of gravity itself but only differences between stars, gas and DM. The constraining power that we lose in marginalisation over an empirical noise model we make up for through the very large number of objects sampled. Galaxies are in fact ideally suited to \emph{relative} tests of gravity between DM and baryons because the former sets their structure and dynamics while the latter are observable.

In future work we will further our analysis of morphological fifth-force signals in hydrodynamical simulations to explore the potential impact of baryonic physics, augment the constraints with additional data and extend our methods to other (e.g. kinematic) galactic modified gravity and dark-sector signals.

\section{Conclusion}
\label{sec:conc}

Modified gravity has been a popular means of explaining cosmic acceleration, ameliorating astrophysical tensions and exploring fundamental physics beyond GR. An archetypal modified gravity model is $f(R)$, where the Ricci scalar in the Einstein--Hilbert action is replaced by a more complex function. Although this is known not to be capable of self-acceleration, it has proven useful as a standard benchmark for gravitational constraints because it typifies the class of models with new scalar fields that induce thin-shell-screened fifth forces. Here we present a test of $f(R)$ gravity by comparing its predictions for galaxy morphology -- separation of stars and gas and warping of stellar disks -- with observational data.

Building on previous work \cite{HIOC, warp}, we construct a galaxy-by-galaxy Bayesian forward model for the morphological signals by combining a reconstruction of the local gravitational and fifth-force fields, models for galaxies' internal structures and a sophisticated empirical model for astrophysical noise. Propagating uncertainties in the input quantities by Monte Carlo sampling and marginalising over the noise parameters, we derive constraints on the strength relative to gravity ($\dG$) and range ($\lambda_C$) of a thin-shell-screened fifth force by comparing its predictions with observations. Our datasets are 15,634 galaxies with HI centroid measurements from ALFALFA and optical measurements from the NSA, and 4,149 edge-on disk galaxies with $r$- and $i$-band images from the NSA, all within 250 Mpc.

Applying slightly more stringent data quality cuts than previously, we find no evidence for a screened fifth force of any $\dG$ in the range $0.3 < \lambda_C/\text{Mpc} < 8$. We thus set the $1\sigma$ constraints shown in Fig.~\ref{fig:constraints}: $\dG<0.8$ at $\lambda_C=0.3$ Mpc up to $\dG<3\times10^{-5}$ at $\lambda_C=8$ Mpc. This excludes the $n=1$ Hu-Sawicki model of $f(R)$ (with $\dG = 1/3$) down to $\lambda_C \simeq 0.38$ Mpc, limiting the background scalar field value to $f_{R0} < 1.4\times10^{-8}$. We also consider other $n$; e.g. for $n=3$ we require $f_{R0} < 10^{-7}$. These constraints -- the strongest to date by over an order of magnitude -- require that practically \emph{all} astrophysical objects be screened, and hence that $f(R)$ and similar models are unable to have any but the slightest effects in astrophysics and cosmology. While we focus on $f(R)$ as a test case, our constraints (with self-screening parameter suitably modified) apply to \emph{all} thin-shell-screened modified gravity models, including the chameleon, symmetron and dilaton mechanisms. We also derive constraints for fifth-force models without screening but with differential coupling to stars, gas and dark matter, finding $\dG \lesssim 10^{-5}$.

Our framework provides a general and intuitive statistical method for forward-modelling galactic (or other) signals under any fundamental physics scenario, with explicit provision for uncertainties in the input quantities and astrophysical contributions to the observables. It also provides a principled way both to combine datasets with constraining power on a given theory and to update constraints as new data becomes available, including automatic recalibration of the noise. These features make it ideally suited to deriving precise and accurate constraints on fundamental physics (gravitational or dark sector) with a range of signals accessed by present and upcoming galaxy surveys, while ensuring robustness to `galaxy formation physics'. We anticipate many more powerful results in the future from studies along these lines.

\bigskip

{\it Acknowledgements:} 

We thank Deaglan Bartlett, Philippe Brax, Clare Burrage, Anne-Christine Davis, Bhuvnesh Jain, Jens Jasche, Justin Khoury, Kazuya Koyama, Guilhem Lavaux, Baojiu Li, John Peacock, Jeremy Sakstein and Hans Winther for useful input and discussions. HD is supported by St John's College, Oxford, and acknowledges financial support from ERC Grant No. 693024 and the Beecroft Trust. PGF is supported by Leverhulme, STFC, BIPAC and the ERC.

\bibliography{ref_F5}

\end{document}